\documentclass[12pt]{elsarticle} 

\usepackage{graphicx}
\usepackage{amssymb}
\usepackage{float}
\usepackage{makecell}
\usepackage{lscape}

\usepackage{hyperref}
\usepackage{longtable}
\usepackage{pdflscape}
\usepackage{multirow, booktabs}
\usepackage{multicol}

\usepackage{tikz} 
\usepackage{graphicx}
\usetikzlibrary{positioning}

\journal{Financial Markets and Portfolio Management.} 

\begin{document}

\begin{frontmatter}


\title{Modeling the yield curve of Burundian bond market by parametric models}



\author[1]{R\'edempteur Ntawiratsa}
\author[2,3]{David Niyukuri}
\author[3,4]{Ir\`ene Irakoze}
\author[2,3]{Menus Nkurunziza}

\address[1]{Department of Economics, Faculty of Economics and Management, University of Burundi, Bujumbura}
\address[2]{Department of Mathematics, Faculty of Science, University of Burundi, Bujumbura, Burundi }
\address[3]{Doctoral School, University of Burundi, Bujumbura, Burundi}

\address[4]{Actuarial Department, Institute of Applied Statistics, University of Burundi, Bubanza, Burundi}

\begin{abstract}

The term structure of interest rates (yield curve) is a critical facet of financial analytics, impacting various investment and risk management decisions. It is used by the central bank to conduct and monitor its monetary policy. That instrument reflects the anticipation of inflation and the risk by investors. The rates reported on yield curve are the cornerstone of valuation of all assets. To provide such tool for Burundi financial market, we collected the auction reports of treasury securities from the website of the Central Bank of Burundi. Then, we computed the zero-coupon rates, and estimated actuarial rates of return by applying the Nelson-Siegel and Svensson models. This paper conducts a rigorous comparative analysis of these two prominent parametric yield curve models and finds that the Nelson-Siegel model is the optimal choice for modeling the Burundian yield curve. The findings contribute to the body of knowledge on yield curve modeling, enhancing its precision and applicability in financial markets. Furthermore, this research holds implications for investment strategies, risk management, second market pricing, financial decision-making, and the forthcoming establishment of the Burundian stock market.

\end{abstract}


\end{frontmatter}








\newpage
\section{Introduction}

With the changes in financial markets in the 1960s, investors became particularly interested in fixed income securities \cite{broze1996estimation}. Indeed, this period saw an increase in interest rates, making these assets more attractive than equities. This upward trend continued in the following decade. In the early 1980s, as a result of US monetary policy, interest rates became more volatile, thus putting bond portfolios at interest rate risk. This is because the value of a bond varies inversely with interest rates and its exposure to interest rate risk increases with its maturity. Naturally, researchers and asset and liability management (ALM) professionals soon became interested in modelling the relationship between bond yields and their maturities. This modelling was limited to sovereign bonds, theoretically immune to default risk, and led to the characterisation of the yield curve or term structure of interest rates.

The term structure of interest rates (yield curve) is a paramount tool for the central bank, the financial intermediaries, and other institutional investors (insurance companies and pension funds). It is used by the central bank to conduct and monitor its monetary policy. The level of future interests and inflation can be predicted by using respectively the spread between long and short term interest rates, and the slope of the yield curve \cite{campbell1991yield}. 

In addition, it guides the central bank in valuation of the sovereign securities it issues for the Treasury (Government). Furthermore, as the sovereign securities are assumed to be risk-free, they serve as reference for valuation of other securities, like corporate bonds, stocks, and other derivatives such as the options, warrants, swaps, forward rates agreements, among others. 


Despite its multiple and crucial applications, the term structure of interest rates is absent in several African countries, including those with bond markets. Consequently, the scientific literature on the term structure of interest rates is far from prolific. Without claiming to be exhaustive, we can cite three bond markets that have benefited from the attention of researchers who have estimated the term structure of interest rates. These are the West African Economic and Monetary Union zone \cite{Souleymanou2021Estimatio, gbongue2019amelioration}, South Africa \citet{aling2012no, moungala2013modelisation}, the CIPRES countries \cite{gbongue2015analyse, gbongue2017proposition}, Kenya \cite{muthoni2015extraction}, Ghana \cite{lartey2018zero}, and Morocco \cite{lekhal2016modelisation}.

Naturally, this void challenges us and leads us to include Burundi among the rare African countries to have this tool with multiple uses both in the evaluation of assets and in financial risk management. Although Burundi still lacks a capital market, it has a fixed income market that is increasingly developing \cite{loi27fevr, Loi29Oct, Decret20Aout} . Prior to 2015, securities issued were largely composed of treasury bills, short-term sovereign securities with maturities of 13, 26 and 52 weeks. In the second half of 2015, in order to cope with the vacuum left by the interruption of budget support by many bilateral and multilateral partners, the government turned to the domestic debt market. An increasing volume of sovereign bonds was issued, on average once (every Wednesday) per week. In 2017, Burundi Central Bank, known as the Bank of the Republic of Burundi (BRB) initiated the secondary market for securities \cite{instr2017}, making them more liquid.

The scale and regularity of bond issues on the primary market \cite{brdetpub}, combined with the increasingly high volume of transactions on the secondary market, make it essential to use the term structure of interest rates to evaluate these assets, which are becoming increasingly important on the balance sheets of banks and insurance companies \cite{brbweb}. To do so, we used primary market data to estimate this structure, given that the secondary market is still in its infancy. In the absence of a standard model applicable to all markets, we set ourselves the objective of proposing a model that is best suited to the Burundian bond market. We then tested two numerical models considered to be adapted to less developed and relatively illiquid bond markets, namely the models of Nelson and Siegel \cite{nelson1985parsimoneous}, and Svensson \cite{dahlquist1996estimating}. 



\subsection{The Burundian financial market}

The Burundian bond market is exclusively made up of sovereign securities (bonds and notes) issued by the Burundian government via its Central Bank. It allows the State to find internal sources of financing while offering investment vehicles mainly to commercial banks and institutional investors, often in need of investment opportunities during this period when the country is facing a double health and economic shock.

Moreover, as in other countries without a capital market, the Burundian financial landscape is dominated by the banking sector. In addition to financing the private sector, they participate in the financing of the state through the purchase of treasury securities, which increasingly dominate the market.

\begin{figure}[H]
    \centering
    \includegraphics[scale=0.8]{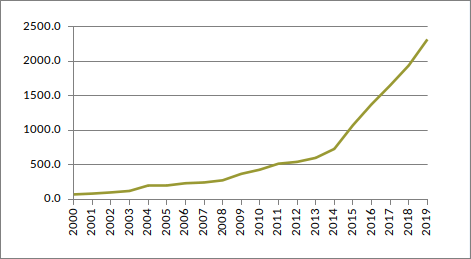}
    \label{fig:dom_pub_debt}
    \caption{Domestic public debt (in billions of BIF).}
\end{figure}

As we can see from supplementary information (\ref{fig:dom_pub_debt}), over a period of twenty years (2000-2019), Burundi's domestic public debt has multiplied by about 34 (overall multiplier), an average annual growth of 20.4\% . This growth accelerated in 2015 following the freezing of funding by the main bilateral and multilateral donors. Thus, from 2015 to 2019, domestic debt increased by a factor of 3.2, or an average annual growth rate of 26.04\%. 

\begin{figure}[H]
    \centering
    \includegraphics[scale=0.8]{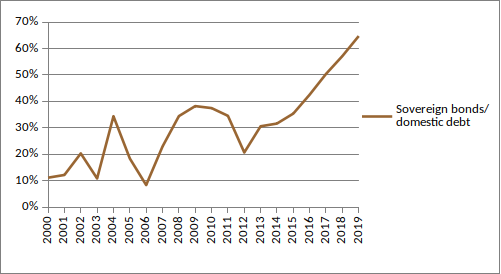}
    \label{fig:share_dom_pub_debt}
    \caption{Share of domestic public debt in treasury bills in the total domestic public debt.}
\end{figure}

From the supplementary information (\ref{fig:share_dom_pub_debt}), we see that the acceleration of domestic debt has been accompanied by the development of the treasury securities market. 

\begin{figure}[H]
    \centering
    \includegraphics[scale=0.8]{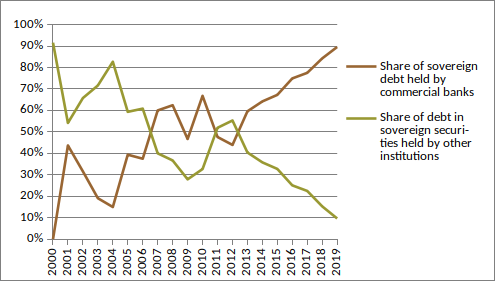}
    \label{fig:evolu_share_com}
    \caption{Evolution of the share of commercial banks in domestic public debt in sovereign securities.}
\end{figure}

There is evidence from the supplementary information (\ref{fig:evolu_share_com}) that, by the growing share of bonds in domestic public debt: while it was only 11\% in 2000, it reached 65\% in 2019 . The proportion of sovereign securities held by commercial banks has increased from 0\% in 2000 to 90\% in 2019. The share held by institutional investors has evolved in the opposite direction so that it is rather marginal in 2019 (10\% against 92\% in 2000).

In response to this development, the Burundi Central Bank has instituted another category of financial intermediary called the Primary Dealer (PD). The PDs are financial intermediaries approved by the central bank to buy Treasury securities in large volumes on the primary market and to sell some or all of them to the public on the secondary market \cite{brbsvt}. They have two main obligations:

\begin{itemize}
    \item to sell at least 10\% of the securities acquired on the primary market
    \item to ensure the stability of Treasury security prices by applying a fluctuation margin of $\pm 10$\%  for maturities of up to 7 years and $\pm 15$\% for terms of more than 7 years in relation to the weighted average price of the previous day's transactions by type of instrument.
\end{itemize}

Moreover, the Burundi central Bank has launched a campaign to promote the secondary market in treasury securities. In addition to a few workshops and media events organised by the department in charge of the money market, the central bank has included in the specifications of the primary dealers the obligation to permanently display at the main entrance, at their counters and on their websites, the two-way quotation (buy-sell) of the different series of Treasury securities they wish to sell or buy.

On the other hand, the PDs have privileges including: (a) exclusive privilege to acquire the securities allocated through the non-competitive method, the amount of which varies between 20-30\% of the overall volume of securities issued; (b) being informed by the Ministry in charge of Finance, at the beginning of each fiscal year, of the forecasted issuance program of Treasury securities.

In Burundi, although PDs have exclusive privilege to acquire securities. From the publicly available data, there is no notice of auction of treasury securities mentions the distribution of the respective amounts between the two auction methods (competitive and non-competitive). Such advantage can distort competition and alter the efficiency of the bond market, which is so much sought after, if we believe the spirit and the letter of the code of ethics and specifications of primary dealers.

The BRB informs the PDs, at least two days before each auction, of the characteristics of the issue in question, in particular the nature of the securities to be issued, their maturity, the amount of the issue, the deadline for the transmission of bids as well as the conditions/methods of auction. The primary dealers are consulted by the Treasury on any change that may be made to the Treasury securities issuance policy. The BRB also consults them on any significant change that may occur in the organisation of the primary and secondary markets for Treasury securities \cite{brbsvt}. This reform should contribute to the efficiency and liquidity of the bond market as a prelude to the long-awaited capital market.

\subsection{Gaps of analytical tools for Burundian financial market}

Despite the progress made in the legal and institutional set-up of this bond market, it lacks a analytic instrument that will allow it to properly value the securities traded. This is the Yield Curve or the Term Structure of Interest Rates. This gap justifies the present work, whose main objective is to fill it. In the following subsection, we describes the conceptual framework of the term structure of interest rates and some methods used to estimate the term structure known also as the yield curve. By providing accurate yield curve estimations, this research will facilitate the evaluation of financial products, the development of trading strategies, and the overall advancement of the Burundian financial bond market.

Moreover, this study has broader implications for the development of the financial sector in Burundi. By improving our understanding of the yield curve dynamics, we contribute to the enhancement of financial products, the formulation of effective trading strategies, and the overall market development. The insights gained from this research will aid policymakers in designing appropriate regulations and policies that foster a sustainable and inclusive financial system in Burundi.

\subsection{African bond market}

The African bond market has gained increasing attention in recent years as a potential source of financing for governments and corporations in the region. Researches have been conducted on the African bond market to understand its growth, challenges, and potential for development \cite{adelegan2009determines, mu2013bond, sy2015trends}.

Studies have examined the historical development of African bond markets, tracing their evolution from limited and underdeveloped markets to more mature and sophisticated ones. Researchers such as O.J. Adelegan and B. Radzewicz-Bak \cite{adelegan2009determines} have analyzed the factors that have contributed to the growth of bond markets in Africa, such as regulatory reforms, government initiatives, and increased investor interest. They have also assessed the impact of external factors, such as global economic conditions and international investor sentiment, on the performance of African bond markets.

Several studies have identified the challenges and constraints faced by African bond markets. Limited liquidity, lack of diversified financial instruments, and shallow investor base are among the key challenges hindering the development of bond markets in the region. Additionally, researchers have examined governance and regulatory issues, market infrastructure, and legal frameworks as potential constraints to market growth (\cite{kahn2005original, oji2015bonds}).

The role of African bond markets in financing economic development has been a focal point of research. Studies have assessed the effectiveness of bond markets in mobilizing domestic savings and funding infrastructure projects. They have also examined the role of bond markets in supporting government borrowing and reducing reliance on foreign debt (\cite{fanta2017equity, kapingura2014causal, berensmann2015developing}).

Researchers have explored the impact of macroeconomic factors on African bond markets. Studies have analyzed the relationship between interest rates, inflation, exchange rates, and bond yields. Understanding these relationships is crucial for investors and policymakers in making informed decisions and managing risks \cite{ahwireng2019macroeconomic}.

Studies on African bond markets have also delved into investor behavior and risk perception. Research has investigated the risk appetite of investors, their preferences for different types of bonds, and the factors influencing their investment decisions in African bond markets \cite{zeitz2022global, gbohoui2023sub}.

In Burundi a stock market has not yet been established  to mobilize savings for investment \cite{nkurunziza2012financial}. The Burundian bond market is exclusively made up of sovereign securities (bonds and notes) issued by the Central Bank. Only the central bank issues treasury bills and bonds to mobilize savings for investment purposes and manage liquidity within the sector. Government bond specialists regularly participate in the issuance of government bonds on the primary market and contribute to secondary market liquidity through the buying and selling of these securities. As in any developing country, the financial system is either inadequate or not as efficient as it should be \cite{patrick1966financial}. 

Government bonds generally have shorter duration compared to corporate bonds. A significant portion of government bonds consists of short-term bonds with maturities of less than one year. T-Bills are frequently utilized for managing liquidity, particularly in the implementation of monetary policies. This can be attributed to factors such as investor preferences, risk tendencies associated with macroeconomic fundamentals, country-specific risks, and the limited depth of the local market. Consequently, the absence of a benchmark yield curve, which could offer pricing indications to prospective issuers of corporate bonds, acts as an obstacle to the growth of the corporate bond market \cite{nyawata2013treasury}.

\subsection{Importance of the yield curve}

Yield curves serve as a valuable tool in understanding how investors collectively perceive the future state of the economy and in guiding financial policy and planning \cite{zaloom2009read}. Interest rates, being a key monetary policy tool for the government, have a significant impact on the shape of the yield curve. The yield curve, which visually depicts the relationship between bond yields, is a crucial indicator for gauging the level and changes in interest rates within an economy. Therefore, it is essential to closely study and model the yield curve.

Moments of volatility in the yield curve require thorough examination, particularly when it flattens or inverts. Under normal circumstances, as time progresses, both risk and borrowing premiums tend to increase. This is because predicting economic conditions becomes more challenging as the time horizon extends beyond a couple of years.

The flattening of the yield curve raises concerns among economic actors as it implies a distorted relationship between risk and time. This phenomenon demands explanation and calls for the development of new strategies for profit and policy-making. The yield curve serves as an important representation of the decentralized and interconnected nature of the modern economy. It provides market participants with a framework to engage in and analyze economic activities\cite{zaloom2009read}. 


According to \cite{muthoni2015extraction}, the yield, also known as the yield to maturity, represents the actual return an investor would receive if a security is held until maturity. It can be defined as the discount rate that equates the present value of all cash flows from the investment to its current price. However, relying on a single discount rate for different time periods presents challenges. This approach assumes that all future coupon payments will be reinvested until maturity, disregarding reinvestment risk and introducing uncertainty over the investment period.

Another limitation of the yield to maturity is its dependence on the cash flow structure, often referred to as the coupon effect. Consequently, the yield-to-maturity of a coupon bond is not an ideal measure of pure time price or the most suitable yield measure for term structure analysis. In contrast, zero-coupon securities eliminate reinvestment risk since they do not have intermediate coupon payments to be reinvested. Additionally, securities with the same maturity exhibit the same spot price, which represents the pure time price in theory, rather than the yield to maturity. Hence, when examining yield curves, it is preferable to utilize the zero-coupon yield curve rather than relying solely on the yield to maturity.

In the absence of a standard model applicable to    all markets, this study sought to propose a model best suited to the Burundian bond market.  Two numerical models considered to be better adapted to less developed and relatively illiquid bond markets, namely the models of Nelson and Siegel
and Svensson were tested.

\section{Materials and methods} 


\subsection{Conceptual framework for the term structure of interest rates}

The value of a coupon bond is equal to the present value of future cash flows consisting of coupons maturing periodically (usually at the end of the year) and the principal repaid at the end of the maturity of the security. Let us take two bonds $A$ and $B$ with respective maturities $n$ and $m$; where $n$ being strictly different from $m$.

Let $P$ be the price of the bond $A$ :
\begin{equation}
    P = \frac{c}{(1+r)}+\frac{c}{(1+r)^2}+\cdots+\frac{c}{(1+r)^n}+\frac{F}{(1+r)^n}
    \label{price_a}
\end{equation}

where $P$ is the price of the bond with maturity $n$, $c$ is the coupon equivalent to the face value of the bond multiplied by the coupon rate, and $r$ the discount rate or actuarial rate of return on the bond.

Let $V$ be the price of bond $B$, it is calculated as follows:

\begin{equation}
    V = \frac{d}{1+g}+\frac{d}{(1+g)^2}+\cdots+\frac{d}{(1+g)^m}+\frac{F}{(1+g)^m}
    \label{price_b}
\end{equation}

where $V$ is the price of the bond with maturity $m$, $d$ is the coupon equivalent to the face value of the bond multiplied by the coupon rate, and $g$ the discount rate or actuarial rate of return on the bond.

Unless there is an inversion of the yield curve, the actuarial yields of bonds issued by the same borrower are a positive function of their maturities. Thus if $m>n\Longrightarrow g>r$. As a result, cash flows (coupons $c$ and $d$) from the same issuer maturing on the same date are discounted at different rates. This constitutes a financial aberration that should be corrected by estimating an identical discount rate for two or more coupons received \textbf{cash inflows} at the same date from the same issuer, i.e. a borrower whose creditworthiness is independent of the maturity of the bonds it has issued $(g = r)$. The most common approach is to consider a coupon bond as a portfolio of zero coupon bonds. Thus, each coupon will be discounted at a rate that varies according to its payment date by the same issuer or seller of the bond (in the case of a secondary market).

\subsection{Estimating the term structure of interest rates}

The interest aroused by the yield curve of fixed-income securities in both academic and professional circles has led to a profusion of estimation methods, some more complex than others and claiming better quality of fit, better smoothing and greater predictive power. However, no model can claim to have all three qualities and is hardly transposable to all bond markets. Hence the existence of several approaches, the relevance of which is subject to the test of the figures from the bond market for which we want to estimate the term structure of interest rates. Far from drawing up an exhaustive list, we can group them into three categories, namely: regression-type models, spline-type models, parametric models, and non-parametric models.

For the regression type models, fixed income yields are expressed as a linear function of several explanatory variables (maturity, coupon rate, etc.). The estimation of the coefficients is classically carried out by minimizing the differences (least squares) between the theoretical returns resulting from the model and those observed on the bond markets.

For splines type models, as the International Bank of Settlements \cite{curves2005technical}(BIS, 2005) has so simply explained, rather than specifying a single functional form for all maturities as in parametric models, spline-based methods fit the yield curve by relying on a piecewise polynomial, the spline function, whose individual segments are smoothly connected at so-called node points. While it is true that the goodness of fit increases with the order of the polynomial, the smoothness and predictive power suffer. Hence, to circumvent this problem, the higher order polynomial is approximated by a sequence of lower order polynomials.

As well as the parametric models are concerned, they have the reputation to fit parsimoniously the yield curve by capturing its main frequently seen shapes: monotonic form, humps at different areas of the curve and the s-shapes. In that class, the Nelson and Siegel model is of primary importance as it is the parent of the Svensson model  \cite{pooter2007examining, diebold2006forecasting}. For this paper, we will focus on the first two models. The Nelson Siegel model is as follows: 

\begin{equation}
    r(T) = \beta_0+\beta_1\frac{\Big(1-e^{-\frac{T}{\tau}}\Big)}{\frac{T}{\tau}}+\beta_2\Big(\frac{1-e^{-\frac{T}{\tau}}}{\frac{T}{\tau}}-e^{-\frac{T}{\tau}}\Big)
    \label{model_ns}
\end{equation}

where $r(T)$ is the rate function to be estimated for a maturity at maturity T expressed in units of years. The model parameters, $\beta_0$ , $\beta_1$ , $\beta_2$ and $\tau$, are parameters to be varied in order to have different curves reflecting specific bond market situations at a given time.

Svensson tried to create a more flexible version by adding an extra term to the existing Nelson-Siegel formula which contained two extra parameters. He thus added an extra term which is nothing but the last term of the Nelson-Siegel equation and replaced the original parameters $\beta_2$ and $\tau$ by $\beta_3$ and $\tau_2$ respectively, thus obtaining the following equation:

\begin{equation}
     r(T)=\beta_0+\beta_1\frac{\Big(1-e^{-\frac{T}{\tau}}\Big)}{\frac{T}{\tau}}+\beta_2\Big(\frac{1-e^{-\frac{T}{\tau}}}{\frac{T}{\tau}}-e^{-\frac{T}{\tau}}\Big)+\beta_3\Big(\frac{1-e^{-\frac{T}{\tau_2}}}{\frac{T}{\tau_2}}-e^{-\frac{T}{\tau_2}}\Big)
         \label{model_sv}
\end{equation}

\subsection{Burundian bond market data}

As we mentioned in the introduction, the Burundian bond market is exclusively made up of sovereign securities (bonds and notes) issued by the Central Bank. In our modelling exercise, we used the daily treasury securities auction results. From the website of the Bank of the Republic of Burundi (Central Bank), we downloaded the auction reports of Treasury securities (bills and bonds). These reports fed into our Excel database mentioning the price, maturity, auction and maturity date, coupon rate and actuarial yield of each security. Initially, we collected this data starting with the August 28, 2014 issue until July 15, 2019. In 2014, there would have been only one issue (28/08/2014) and two in 2015 (28/01/2015 and 09/12/2015). It was only in 2016 that the frequency of bond auctions increased to the weekly level (every Wednesday of the week) generally observed from 2017 onwards. 

In total, we collected raw data for 112 bond and bill auction dates with maturities ranging from 13 weeks (0.25 year) to eight years. This size was considerably reduced because the Matlab software we used required us to process only securities with a full range of maturities. For example, we had to avoid bonds with a maturity of more than 5 years, as well as some dates that missed certain maturities. In the end, the sample size fell to 30 auction dates, each comprising seven securities, including three treasury bills (13, 26 and 52 weeks) and four bonds (2, 3, 4 and 5 year maturities). As a result, the estimation of our curve was based on 210 Burundian treasury securities.

We had two data-sets to explore the yield curve modelling, one data-set of complete maturities which were issued at 23 different days by the Central Bank. That means, for those dates, the Central Bank issued maturities for all years from one to the maximum and all were taken. The second data-set was from 67 different days, but which had missing maturities. The latter were imputed by filling in missing values with previous values. That means, if for example we had maturity 2 and 5, those missing were imputed. For each missing maturity we had to impute the price, the coupon rates, and the yield. The choice for that imputation method was supported by internal results when comparing missing data imputation methods for bond market in Burundi.

\subsection{Model calibration: parameters estimation and optimization}

In order to estimate Svesson and Nelson-Siegel parameters models, as some of the parameters have economical meaning which are bounded between minimum and maximum values, the calibration becomes an optimization problem. To find the best fit parameters estimates, we used the Global Optimization by Differential Evolution (DE), using the \textit{DEoptim} R package  \cite{mullen2011deoptim}. The DE is a type of genetic algorithm developed by Rainer Storn and Kenneth Price in the 1990s \cite{storn1997differential}. It is reported to be well-suited to find the global optimum of a real-valued function of real-valued parameters, and does not require that the function be either continuous or differentiable. Thus, for our yield curve models given by the equations above (\ref{model_ns} and \ref{model_sv}).

\subsection{Statistical analysis}

After calibration, we took the parameter estimates and run the models (Svesson and Nelson-Siegel), and compare the results to the zero coupon rates, by computing root mean square error (RMSE). For both models and each data-set, we had a set of error values from Svesson and Nelson-Siegel models. 
Having the imputed data with the sample size of 67 observations which does not follow a normal distribution, we use straightly the paired Wilcoxon Rank Sum test \cite{whitley2002statistics} to assess if error medians from the two models are equal. From the median values of the error values, and the Wilcoxon Rank Sum test, we get to know the best fit yield-curve model. For the original data with 23 observations which follows a normal distribution, a paired t-test to compare the means of the two models Nelson-Siegel and Svensson.

\section{Results}

\subsection{Parameter estimation}

Per each bonds issuance date, we calibrated the Svensson and Nelson-Siegel models using the bonds data of the day. Thus, when using complete data (from the 23 days), we got the following parameter estimates summarized in the following Table \ref{tab:params_complete}, and using imputed data (from the 67 days), we got the following parameter estimates summarized in the following Table \ref{tab:params_imputed}.


\begin{table}[ht] 
\centering
\caption{Summary statistics for parameters values form the 23 days original data}

 {\small
    \resizebox{\textwidth}{!}{
    \begin{tabular}{p{4cm} p{2cm} p{2cm} p{2cm}  p{2cm} p{2cm} p{2cm}} 
    \hline
    
    \textbf{Model}  &  \textbf{Parameter}  & \textbf{Median} & \textbf{Mean} & \textbf{Q.1} & \textbf{Q.3} & \textbf{Variance} \\ \hline

\multirow{4}{*}{Nelson-Siegel} 
 & $\beta_0$ & 8.810  &10.819 &  6.914 & 14.275 &16.4461 \\
 & $\beta_1$ & -5.985 & -8.315&  -13.269  & -3.103   & 27.9443 \\
 & $\beta_2$ &  -1.932 &-4.125 & -7.641 & -1.477 & 10.6141 \\
 & $\tau$ & 1.0296  &1.2193 &1.0002   &1.3517 &  0.1129 \\
\hline

 \multirow{6}{*}{Svensson} 
 & $\beta_0$ &  8.796  &10.806  &  6.912 & 14.306 & 16.3325\\
 & $\beta_1$ & -6.306 & -8.449&  -13.280 & -3.279 & 27.2011\\
 & $\beta_2$ &-1.932&-4.063 & -7.461  & -1.511 & 10.1219 \\
 & $\beta_3$ & 0.9287 & 0.6721 & 0.2498 & 0.9866& 0.1684 \\
 & $\tau_1$ &  1.0182 & 1.2195 & 1.0006 & 1.3549 &0.1090  \\
 & $\tau_2$ & 0.7580 & 0.6029 & 0.2205&0.9578 & 0.1331 \\
 \hline
 \end{tabular} } }
    \label{tab:params_complete}
\end{table}


Table \ref{tab:params_complete} provides the summary statistics for the Nelson-Siegel and Svensson models for the complete data. Whereas Table \ref{tab:params_imputed} provides the summary statistics for the Nelson-Siegel and Svensson models for imputed data.

Looking at the \textbf{Nelson-Siegel Model} from Table \ref{tab:params_complete}, we can see that for the parameters $\beta_0$, $\beta_1$, $\beta_2$, and $\tau$, we have the following observations. 

The median values are within a reasonable range for these parameters, suggesting that they are centered. However, the mean values are slightly higher, indicating that the distributions might be slightly right-skewed. This skewness could be seen in the positive difference between the mean and median. The variances for these parameters are relatively high, indicating a wide spread of data points around the mean or median. This suggests some level of data dispersion, which might not align perfectly with a normal distribution. The  parameter $\tau$ appears to have a relatively narrow spread, with a small variance. This suggests that the values of $\tau$ are somewhat concentrated around the median.

For the \textbf{Svensson Model}, similarly to the Nelson-Siegel model for the parameters $\beta_0, ~\beta_1, ~\beta_2, ~\beta_3, ~\tau_1$, and $\tau_2$: the median values are within a reasonable range for these parameters, but the mean values are slightly higher, indicating potential right-skewness. The variances for these parameters are again relatively high, indicating variability in data points.

The  parameters $\tau_1$, and $\tau_2$ also have relatively small variances, suggesting that the values are concentrated around the median.



\begin{table}[ht] 
\centering
\caption{Summary statistics for parameters values from the 67 days imputed data}

 {\small
    \resizebox{\textwidth}{!}{
    \begin{tabular}{p{4cm} p{2cm} p{2cm} p{2cm}  p{2cm} p{2cm} p{2cm}} 
    \hline
    
    \textbf{Model}  &  \textbf{Parameter}  & \textbf{Median} & \textbf{Mean} & \textbf{Q.1} & \textbf{Q.3} & \textbf{Variance} \\ \hline

\multirow{4}{*}{Nelson-Siegel} 
 & $\beta_0$ & 9.578  &10.494 &  8.361 & 12.671 & 10.2151 \\
 & $\beta_1$ & -5.841 & -7.536&  -9.317  & -4.192  & 18.0525 \\
 & $\beta_2$ &  -2.932 & -4.014 & -5.888 & -2.039 & 6.3197 \\
 & $\tau$ &  1.3016  & 1.4369 & 1.0013  & 1.7928 &  0.2655 \\
\hline

 \multirow{6}{*}{Svensson} 
 & $\beta_0$ &  9.572  & 10.493 &  8.360 & 12.612 & 10.1017\\
 & $\beta_1$ & -6.077 & -7.677 &  -9.671 & -4.191 & 17.5200\\
 & $\beta_2$ &-3.044& -4.076 & -5.719  & -2.096 & 6.0687 \\
 & $\beta_3$ & 0.9310 & 0.7554 & 0.6400 & 0.9832& 0.1136 \\
 & $\tau_1$ &  1.3533 & 1.4528 & 1.0057 & 1.8041 &0.2821 \\
 & $\tau_2$ & 0.7782 & 0.6365  & 0.2777 &0.9515 & 0.1209 \\

 \hline

    \end{tabular} } }
    \label{tab:params_imputed}

\end{table}

From the summary statistics in Table \ref{tab:params_imputed} for the Nelson-Siegel and Svensson models based on the 67 days imputed data, we have the observations below.

For the \textbf{Nelson-Siegel Model} looking at the parameters $\beta_0$, $\beta_1$, $\beta_2$, and $\tau$ we can see that the median values are within a reasonable range for these parameters, suggesting that they are centered. However, the mean values are slightly higher than the medians, indicating potential right-skewness. This skewness could be seen in the positive difference between the mean and median. The variances for these parameters are relatively moderate compared to the previous data, indicating less variability in data points around the mean or median. This suggests that the imputed data might have reduced the spread of the data. The parameter $\tau$ appears to have a relatively narrow spread, with a small variance. This suggests that the values of $\tau$ are somewhat concentrated around the median.

For the \textbf{Svensson Model}, the median values  of the parameters $\beta_0$, $\beta_1$, $\beta_2$, $\beta_3$, $\tau_1$, and $\tau_2$ are within a reasonable range for them, but the mean values are slightly higher, indicating potential right-skewness. The variances for these parameters are moderate, similar to the Nelson-Siegel model, indicating some variability in data points.
$\tau_1$ and $\tau_2$ also have moderate variances, suggesting that the values are moderately concentrated around the median.

\begin{figure}[H]
    \begin{center}
        \includegraphics[scale=0.8]{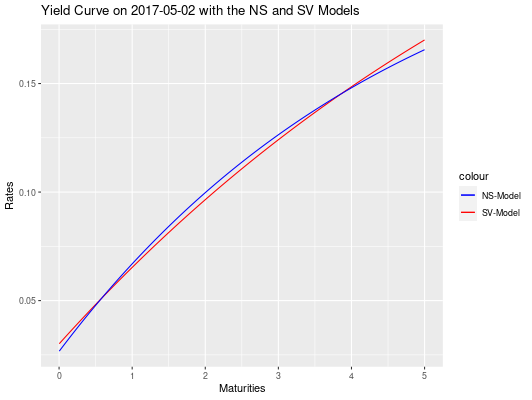}
    \end{center}
     \caption{Models comparison}
    \label{comparison}
\end{figure}

\subsection{Best fit yield-curve model}

From model parameters above, and error values summarized in tables \ref{tab:error_imputed} and \ref{tab:error_original} , it is difficult to distinguish between the two models, as their respective curves are so close. This is illustrated in figure \ref{comparison}, where the observed zero-coupon rate curve is almost identical to those from the Nelson-Siegel and Svensson models.

\begin{table}[ht] 
\centering
\caption{Summary statistics for error values with the 67 days imputed bonds data} 

    \begin{tabular}{p{2.5cm}  p{2cm} p{2cm}  p{1.5cm} p{1.5cm} p{1.5cm}} 
    \hline
    
    \textbf{Model}    & \textbf{Median} & \textbf{Mean} & \textbf{Q.1} & \textbf{Q.3} & \textbf{Variance} \\ \hline
{Nelson-Siegel}& 
   0.7129 & 0.8445 & 0.5364 &  0.9920 & 0.2090 \\
\hline

{Svensson} &
0.7274  & 0.8009 &  0.5037 & 0.9492 & 0.1945\\
 
 \hline
   \end{tabular} 
    \label{tab:error_imputed}

\end{table} 

Table \ref{tab:error_imputed} provides the summary statistics for error values with the 67 days imputed bonds data for both the Nelson-Siegel and Svensson models and we can see that.

For the \textbf{Nelson-Siegel Model}, the median error value is approximately 0.71288, while the mean error value is slightly higher at around 0.84447. This suggests that the distribution of error values may be slightly right-skewed, as the mean is greater than the median. The first quartile ($Q_1$) is approximately 0.53637, and the third quartile ($Q_3$) is approximately 0.99204. These quartiles indicate the spread of the data. The interquartile range (IQR) is relatively wide, suggesting that the error values have a significant spread.

For the \textbf{Svensson Model}, the median error value is approximately 0.72740, and the mean error value is approximately 0.80088. As with the Nelson-Siegel model, the mean is slightly higher than the median, indicating potential right-skewness. The first quartile ($Q_1$) is approximately 0.50366, and the third quartile ($Q_3$) is approximately 0.94915. Similar to the Nelson-Siegel model, the IQR is relatively wide, indicating a significant spread in error values.

\begin{table}[ht] 
\centering
\caption{Summary statistics for error values with the 23 days original bonds data}

    \begin{tabular}{p{2.5cm}  p{2cm} p{2cm}  p{1.5cm} p{1.5cm} p{1.5cm}} 
    \hline
    
    \textbf{Model}    & \textbf{Median} & \textbf{Mean} & \textbf{Q.1} & \textbf{Q.3} & \textbf{Variance}\\ \hline
{Nelson-Siegel}& 
   0.8897& 0.8455& 0.3717& 1.0555 & 0.2822\\
\hline

{Svensson} &
 0.8989& 0.9006& 0.4951& 1.0970 & 0.3130\\
 
 \hline
   \end{tabular} 
    \label{tab:error_original}

\end{table} 

Table \ref{tab:error_original} shows the summary statistics for error values with the 23 days original bonds data for both the Nelson-Siegel and Svensson models.

For \textbf{Nelson-Siegel Model}, the median error value is approximately 0.88973, and the mean error value is approximately 0.84548. In this case, the median is slightly higher than the mean, suggesting potential left-skewness, although the difference is not substantial. The first quartile  is approximately 0.37172, and the third quartile  is approximately 1.05551. The interquartile range  is relatively wide, indicating a significant spread in error values.

For the \textbf{Svensson Model}, the median error value is approximately 0.89887, and the mean error value is approximately 0.90064. The median and mean are very close in this case, suggesting a relatively symmetrical distribution. The first quartile is approximately 0.49509, and the third quartile 
is approximately 1.09704. Similar to the Nelson-Siegel model, the IQR is wide, indicating a notable spread in error values.

In the Nelson-Siegel model, the slight difference between the median and mean may indicate a minor skewness in the distribution of error values, potentially leaning towards the left. However, in the Svensson model, the median and mean are very close, suggesting a more symmetric distribution of errors.

\begin{table}[H]
\caption{Results for the Shapiro-Wilk normality test}
    \centering
    \begin{tabular}{c|c|c}
    \hline
    \textbf{Model} & \textbf{Data }& \textbf{p-value }\\
    \hline
   Nelson-Siegel & Original&0.1363\\
    \hline
     Nelson-Siegel & Imputed &$3.15\times10^{-5}$\\
      \hline
   Svensson& Original &0.0579\\
    \hline
   Svensson& Imputed &$3.6\times10^{-4}$\\
    \hline
    \end{tabular}
    \label{tab:normtest}
\end{table}

By conducting the Shapiro-Wilk test for normality, as presented in Table \ref{tab:normtest}, it becomes evident that the imputed data from both models do not follow a normal distribution. In contrast, the original data conforms to a normal distribution. Given the normality assumption and the ample sample size of the original data, a t-test has been employed to compare the two models, Nelson-Siegel and Svensson.

\begin{table}[H]
 \caption{Comparison of Nelson-Siegel and Svensson models with the two data sets}
    \centering
    \begin{tabular}{c|c|c}
    \hline
      \textbf{Data}& \textbf{Test}   & \textbf{p-value }\\
      \hline
      Original   &t-test&$2.678\times 10^{-3}$ \\
      \hline
    Imputed  &Wilcoxon sum rank &$3.867\times 10^{-7}$\\
    \hline
    \end{tabular}
    \label{tab:t-test_wilcoxon}
\end{table}

The results, displayed in Table \ref{tab:t-test_wilcoxon} above, unequivocally indicate a significant difference between these two models at a 95\% confidence level. Furthermore, the outcomes of the Wilcoxon Rank Sum test, when considering the imputed data, also demonstrate a substantial distinction between the Nelson-Siegel and Svensson models, again with a 95\% confidence level.

\section{Discussion and Conclusion}

Looking at the summary statistics in Table \ref{tab:params_complete} for the Nelson-Siegel and Svensson models based on the 23 days original data, we have the observations below. In both models, the differences between the median and mean, as well as the relatively high variances, suggest that the data for these parameters may not perfectly follow a normal distribution. There may be some skewness or heavy-tailed behavior in the distributions. 

Looking at the summary statistics in Table \ref{tab:params_imputed} for the Nelson-Siegel and Svensson models based on the 67 days imputed data, we have the observations below. In both models, as with the previous data, the differences between the median and mean values, as well as the variances, suggest that the data for these parameters may not perfectly follow a normal distribution. There may still be some skewness or deviations from normality in the distributions.

From the summary statistics for error values  in Table \ref{tab:error_original}, we can see that in both models, the differences between the median and mean error values, as well as the wide IQR, suggest that the error values may not perfectly follow a normal distribution. The potential right-skewness in both cases could imply that there are some larger positive errors in the dataset.

From the summary statistics for error values  in Table \ref{tab:error_imputed}, we can see that in both models, the differences between the median and mean error values, as well as the wide IQR, suggest that the error values may not perfectly follow a normal distribution. The potential right-skewness in both cases could imply that there are some larger positive errors in the dataset.

If we look at the medians of the error values, we can see that in both situation with complete and imputed data, the Nelson-Siegel model was the best one. From Table \ref{tab:error_imputed} and \ref{tab:error_original}, we can see that as the data increase, the variance of the error values decreases. This is supported by the quantiles values which decrease as well. The mean error value for the Nelson-Siegel model in both situations of the data has a relatively stable value (0.84), which is the small value compared to what we get with Svensson model when we have small data.

We also know that in the normality test, if the sample size is small, the power is not guaranteed \cite{anderson2017sample}. Small sample size may likely pass the normality test, thus, we claim that, the normality of error values when we have small original data is not guaranteed. We can see that with big sample size (imputed data), we did not have normal distribution of the error values. 

In general, we conclude that the choice between these two models (Nelson-Siegel and Svensson) has a substantial influence on the outcomes. A significant difference prompt further investigation into why one model may outperform the other. But, the quality assurance for the data is predominant for the precision and accuracy. Based on the results of the comparison tests and the summary statistics, including the medians and variances, we can conclude that the Nelson-Siegel model is the best fit for the Burundian yield curve. Our result substantiates the conclusion made by Zoricic (2013)\cite{zoricic2013parametric} in their study, which suggests that the Svensson model appears to be excessively parameterized in the context of an illiquid and underdeveloped financial market.
\section*{Declarations}


\begin{itemize}
\item \textbf{Funding :} No funding was received to assist
with the preparation of this manuscript.
\item \textbf{Conflict of interest:} The authors have no competing interests to declare that are relevant to the content of this article.
\item \textbf{Ethical statement: }This manuscript is exclusively submitted to this journal. The research presented within this manuscript addresses the critical domain of handling missing data. The research conducted rigorously and transparently by the authors, stands as an ethical pursuit aiming to enhance precision and applicability in financial markets. This research endeavors, devoid of bias and motivated by advancing knowledge, adhere to the ethical principles of academic inquiry. The ethical integrity in data collection, model application, and analysis stands as a testament to the commitment to fostering informed financial decision-making and contributing responsibly to the development of Burundian financial landscape. The content presented in this manuscript is the original work of the authors.
\end{itemize}

\bibliographystyle{elsarticle-num-names}
\bibliography{yield_curve_financial_market.bib}
\section*{Appendix}





\section*{Supplementary information 1: Domestic public debt}

\section*{Supplementary information 2: Share of domestic public debt in treasury bills in the total domestic public debt}

\section*{\textbf{Supplementary information 3: } Evolution of the share of commercial banks in domestic public debt in sovereign securities}

\newpage

\begin{landscape}

\section*{\textbf{Supplementay information 4: } Economic meaning of Model Parameters}

\begin{center}
\begin{table}[H]
\caption{Model Parameters and their economic meaning}
\begin{tabular}{c|c|c}
\hline
\textbf{Parameter}&\textbf{Description}& \textbf{References}\\
\hline
$\beta_0$ &\makecell[l]{ The long term interest rate}&\cite{patrick2015nelson},\cite{huang2021fitting}.\\
\hline
$\beta_1$& \makecell[l]{related to the slope of the yield curve.It signifies how the yield curve\\ changes as you move from short-term to long-term maturities.\\ Its higher absolute value indicates a steeper slope.}&\cite{bolder1999yield},\cite{huang2021fitting}.\\
\hline
$\beta_0+\beta_1$& \makecell[l]{The instantaneous short term interest rate}&\cite{patrick2015nelson}\\
\hline
$\beta_2$&  \makecell[l]{introduces curvature to the yield curve.\\ It determines the curvature's magnitude, which\\ can make the curve either upward or downward sloping.}&\cite{bolder1999yield},\cite{bank2014zero},\cite{huang2021fitting}.\\
\hline
$\beta_3$& \makecell[l]{ provides an extra degree of freedom\\ for modeling the curvature of the yield curve.\\It determines the magnitude and direction
of the curvature.}&\cite{bolder1999yield},\cite{bank2014zero}.\\
\hline
$\tau$& \makecell[l]{ known as the exponential decay factor.\\ It controls the speed at which yields converge\\ to the long-term rate as the time to maturity increases.\\ Smaller values of $\tau$ result in faster convergence.}&\cite{bolder1999yield}, \cite{bank2014zero}.\\
\hline
$\tau_1$& \makecell[l]{determines the speed at which yields converge\\ to the long-term rate for short-term maturities}&\cite{bolder1999yield}, \cite{bank2014zero}.\\
\hline
$\tau_2$& \makecell[l]{controls the speed of convergence for long-term maturities}&\cite{bolder1999yield}, \cite{bank2014zero}.\\
\hline
\end{tabular}
\end{table}
\end{center}
\end{landscape}






\end{document}